\documentclass[12pt,fleqn]{article}
\usepackage{amsmath}
\usepackage{graphicx}
\usepackage{multirow}

\begin{document}
\bibliographystyle{prsty}
\title{Conclusive quantum-state transfer with a single randomly coupled spin chain}
\author{Jiankui He\footnote{{\it E-mail addresses:} cam@mail.ustc.edu.cn} ,
 Qing Chen, Lei Ding, Shao-Long Wan
  \\
\small\it{Department of Modern Physics, University of Science and Technology of China,} \\
\small\it{Hefei 230026, PR China}\\
 }

\maketitle

\begin{abstract}
\baselineskip20pt

We studied the quantum state transfer in randomly coupled spin
chains. By using local memories storing the information and dividing
the task into transfer portion and decoding portion, conclusive
transfer was ingeniously achieved with just one single spin chain.
In our scheme, the probability of successful transfer can be made
arbitrary close to unity. Especially, our scheme is a good protocol
to decode information from memories without adding another spin
chain. Compared with Time-reversed protocol, the average decoding
time is much less in our scheme.

\end{abstract}

PACS: 03.67.Hk, 05.60.Gg, 75,10.Pq  \baselineskip20pt

\newpage
\section{Introduction}
The task of Quantum state transfer(QST) is to transfer a quantum
state  from the sender(Alice) to the receiver(Bob). In 2003, Bose
\cite{r1} presented a scheme to transfer states using unmodulated
spin chains. After that, much effort has been devoted to developing
this scheme [2-19]. In this scheme, we encode the state at one end
of the spin chain and wait for a specific amount of time, then we
get the state at another end of the chain. All the operations and
measurements are performed at the sending and receiving qubits, but
the spin chain connecting them is under no operation.

A simple example of this scheme is a one-dimension spin chain of
XY model interaction with equal nearest-neighbor coupling.
However, perfect QST can be achieved only when the number of spins
is less than four \cite{r1,r3}. Based on the Cartesian Product
method of the graph theory, Christandl {\it et al} \cite{r3}
extended the case of two or three spins to spin networks which can
also get perfect QST. Considering the symmetry of the
Hamiltonian(i.e., mirror symmetry, translational symmetry),
another approach is to use the Hamiltonians with engineered
couplings. Recently, more attention was payed to get perfect QST
with randomly coupled quantum spin chains. A ``dual-rail" encoding
using two parallel quantum channels \cite{r10,r11} is suggested to
get conclusive and arbitrary perfect QST. Here, the word
``conclusive" means whether the state is transferred successfully
can be determined by measuring one qubit of Bob's without
destroying the information. Another method is to use local
memories \cite{r9}, by swapping the qubit of Bob to the memory at
equal time intervals, the whole information will be stored in the
memories and can be decoded by Time-reversed protocol.

The aim of this paper is to replace the ``dual-rail" scheme
\cite{r10} with a single spin chain and reach perfect and
arbitrary quantum state transfer. Our scheme  also provides a
convenient and rapid decoding protocol for ref.\cite{r9}.  The
first step of our scheme is to transfer the information to a
collection of memories by continually swapping between Bob and the
memories, and the second step is decoding the information from the
memories.

\section{Our scheme for perfect QST}
%%%%%%%%%%%%%%%%%%%%%%%%%%%%%%%%%%%%%
\begin{figure}
  \centering
 \includegraphics[scale=0.7]{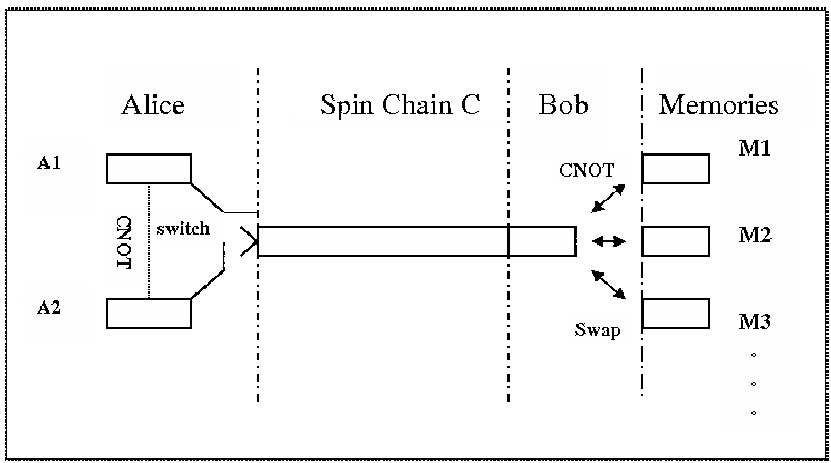} \hspace{0.5cm}
  \caption{Perfect quantum state transfer in one spin chain using local memories}
   \label{fangan}
\end{figure}
%%%%%%%%%%%%%%%%%%%%%%%%%%%%%%%%%%%%%%%%%%%%%5
Our scheme is divided into four parts: A(Alice), B(Bob), C(other
parts in the spin chain), M(memories). See Fig.1. A contains two
identical portions $A_1$ and $A_2$ which have the same number of
spins . When $N_A>1$, it is convenient to transfer multi-qubit
states and entangled states. There are switches connecting the
first spin of C and $A_1$,$A_2$ respectively. Although it is
difficult to continually switch ``on" and ``off" in the spin
channel, we use only limited switches at Alice's site, this will
not increase the complex of the system obviously. The Hamiltonian
of spin chain $A_1+C+B$ and $A_2+C+B$ are identical, and it can be
randomly coupled Heisenberg model or XY model. There is a NOT gate
\cite{chang} between $A_1$ and $A_2$, controlled by $A_1$ being
zero. B contains $N_B$ spins, in our scheme, we will perform local
operations and measurements at Bob's site. M is a series of
memories: $M_1,M_2,M_3\cdots$, and every memory has the same spins
as Bob's. We call it ``memory" because Bob can perform a swap
between his spins and every memory and store the state of Bob in
the memories. Besides, Bob can apply a CNOT gate between his qubit
and memories which will be discussed in decoding portion . The
number of memories is decided by the fidelity we required and the
ability of transfer of the spin chain. C is a spin chain with
$N_C$ spins, and the coupling strength can be arbitrary. We do not
perform any operation at C in our scheme. Therefore, the total
length of chain $A_1+C+B$ (and $A_2+C+B$ ) is $N=N_A+N_B+N_C$. In
this paper, we just consider $N_A=N_B=1$ .

%The key ideal is transfer the quantum state to the memories by
%several steps, and make sure the state stored in a certain memory
%by measurements.

\section{Perfect state transfer to memories}
The Hamiltonian we considered  is  given by \begin{equation}
H_G=\sum_{(i,j)} J_{ij}\Bigl[  \sigma_i^x \sigma_j^x+\sigma_i^y
\sigma_j^y ] +\sum_{i }B_i \sigma_{i}^z \label{hhh}
 \end{equation}
Where $J_{ij}$ is the coupling strength between the \textit{i}th and
the \textit{j}th spin. $\overrightarrow{\sigma}^i =(\sigma ^i_x ,
\sigma ^i_y , \sigma ^i_z )$ is the Pauli matrix of the \textit{i}th
spin. $B_i > 0$ are the static magnetic fields. From Eq.(\ref{hhh}),
it is easy to get $[H,\sigma^z_{tot}]=0$, where
$\sigma^z_{tot}=\sum_{i} \sigma ^z_i$. So, $\sigma^z_{tot}$ is
conserved.

 First, $A_1$ and $A_2$ is separated from  the first
qubit of C. the whole system is cooled to its ground state
$|0\rangle=|00\cdots\cdots0\rangle$, in which $|0\rangle$ denotes
the spin down state(i.e.,the spin aligned -z). $H|0\rangle =E_g
|0\rangle$, and we set the ground energy $E_g=0$. And
$|n\rangle=|00\cdots010\cdots0\rangle$(where $ n=1,2\cdots,N$)
denotes the \textit{n}th spin up, and other spins down.
%Therefore, the states $\{{|j\rangle},j=1,2,\cdots,N\}$ make up a collection of complete base
%vectors.
The state that is transferrd is $|\psi\rangle = \alpha |1\rangle +
\beta |0\rangle$( where $\alpha,\beta \in C$ and
$|\alpha|^2+|\beta|^2=1 $), Alice encodes this state at $A_1$,
this state of the whole system is  \begin{equation} |\psi
0000\rangle_{A_1 A_2 C B M}=|\psi\rangle_{A_1}\otimes
|0\rangle_{A_2} \otimes
|0\rangle_{C}\otimes|0\rangle_{B}\otimes|0\rangle_{M}
\label{chutai}
\end{equation} in which $|0\rangle_{M}$ is a notation of the state
of $k$ memories, $|0\rangle_{M}=|0\rangle_{M_1}\otimes
|0\rangle_{M_2}\otimes\cdots\otimes |0\rangle_{M_k}$. We define
\begin{equation}
|j\rangle_M=|0\rangle_{M_1}\otimes|0\rangle_{M_2}\otimes\cdots\otimes|1\rangle_{M_j}\otimes\cdots|0\rangle_{M_k}
\end{equation}
which denotes the spin of the $j$th memory up and others down.

To make sure that the information is not destroyed by the
measurement, we apply a CNOT gate ($|0\rangle \langle 0 |\otimes
\sigma_x +|1\rangle \langle 1 | \otimes I$) at $A_2$, So $A_1,A_2$
is entangled, the state is
\begin{equation} |\Psi(t_0=0)\rangle=(\alpha|10\rangle
+\beta|01\rangle)_{A_1,A_2} \otimes
|0\rangle_{C}\otimes|0\rangle_{B}\otimes|0\rangle_{M} \label{cnot}
\end{equation}
Then, we switch ``on" between $A_2$ and chain C,  use the notation
$\widetilde{C}$ to represent the whole chain $A_2+C+B$, and label
the spins of $\widetilde{C}$ as $n=1,2,\cdots,N$(the 1st spin is
$A_2$, the $\textit{N}$th spin is Bob).  Generally, we suppose
that the time spent on the switch is much shorter than the system
dynamics. According to the dynamics dictated by Eq.(\ref{hhh}),
the state Eq.(\ref{cnot}) will evolve. At time $t_1=\tau_1$, the
state is given as
 \begin{eqnarray}
 |\Psi'(t_1)\rangle &=& U(\tau_1)|\Psi(t_0=0)\rangle \nonumber\\
&=& \alpha |100\rangle_{A_1 \widetilde{C} M} + \beta \sum^{N}_{n=1}
f_{n}^{(1)}|0n0\rangle_{A_1 \widetilde{C} M} \label{t1}
\end{eqnarray} in which $U(\tau_1)=e^{-i H \tau_1}$ and $f^{(1)}_n =
_{A_1 \widetilde{C} M}\langle 0n0 |e^{-i H \tau_1}
|010\rangle_{A_1 \widetilde{C} M}$ . Then, we perform a
swap($S_1$) between the first memory $M_1$ and Bob'qubit, to
exchange the states. The state becomes
 \begin{eqnarray}|\Psi(t_1)\rangle&=&
S_1|\Psi'(t_1)\rangle \nonumber\\
&=& \alpha |100\rangle_{A_1
\widetilde{C} M} + \beta \sum^{N-1}_{n=1} f_{n}^{(1)}|0n0\rangle_{A_1
\widetilde{C} M} +
\beta f_{N}^{(1)}|001\rangle_{A_1
\widetilde{C} M}\nonumber\\
\label{t2} \end{eqnarray}  $M_1$ has stored some information of
the state and Bob is at the  state $| 0\rangle$. Then the whole
spin chain continues to evolve, at time $t_2=\tau_1 +\tau_2$,
perform a swap $S_2$ between Bob's qubit and $M_2$, the state is
\begin{equation}
|\Psi(t_2)\rangle =S_2 U(\tau_2)|\Psi(t_1)\rangle
\end{equation}
Repeat this operation at time $t_3=t_2+\tau_3,
t_4=t_3+\tau_4,\cdots.$ After the \textit{j}th step, the state of
the system can be described as
\begin{eqnarray}
|\Psi (t_j)\rangle &=&S_j U(\tau_j)S_{j-1}U(\tau_{j-1})\cdots S_1
U(\tau_1)|\Psi(t_0)\rangle\nonumber\\
&=&\alpha |100\rangle_{A_1 \widetilde{C} M} + \beta \sum^{N-1}_{n=1} f^{(j)}_{n}
|0n0\rangle_{A_1 \widetilde{C} M} +\beta \sum _{l=1} ^j f^{(l)}_N |00l\rangle_{A_1
\widetilde{C} M} \nonumber\\
 \label{tjj}
\end{eqnarray}
 In which, \begin{equation}f_n^{(l)}=\sum_{m_1=1}^{N-1}  \sum_{m_2=1}^{N-1} \cdots \sum_{m_{l-1}=1}^{N-1}
\{ f_{m_1,1}(\tau_1) f_{m_2,m_1}(\tau_2) \cdots f_{n,m_{l-1}}(\tau_l)\},\end{equation}
and
\begin{equation}f_{m,n}(t)=
_{A_1\widetilde{C} M}\langle 0m0|U(t)|0n0\rangle_{A_1
\widetilde{C} M}\end{equation} with $l=1,2,\cdots,j$. When the
number of steps $j\rightarrow\infty$, all the excitation state in
spin chain $A_2+C+B$  will be transferred to the memories, and the
spin chain is at its ground state \cite{r9}. \begin{equation}
\lim_{j\rightarrow\infty}|\Psi(t_j)\rangle =\alpha
|100\rangle_{A_1 \widetilde{C} M} +\sum _l \beta f^{(l)}_N
|00l\rangle_{A_1 \widetilde{C} M}\end{equation} So, we have
already stored all the information in the memories.

\section{Decoding information from memories}
After the $j$th step transfer, what we should do is decoding the
state $|\psi\rangle$ from the memories. Time-reversed protocol was
introduced in Ref.\cite{r9} to recover the information. This
protocol requires an another identical spin chain for decoding and
spends the same amount of time as transfer portion. Moreover, it
is not conclusive transfer. Here, we present a new decoding
protocol which can achieve conclusive transfer and spend much less
time than Time-reversed protocol.

 Cool the spin chain to the ground state $|0\rangle_{\widetilde{C}}$, then
switch ``off" between $A_2$ and C, and switch ``on" between $A_1$
and C. $A_1+C+ B$ compose a new spin chain identical to $A_2+C+B$
and we label it as $\widehat{C}$. Let the spin chain evolve
freely, at time $t_1=\tau_1$, the state is given by
\begin{eqnarray}
 |\widehat{\Psi}(t_1)\rangle
&=&U(\tau_1)|\Psi(t_j)\rangle \nonumber\\
&=&  \alpha\sum^{N}_{n=1} f_{n}^{(1)}|0n0\rangle_{A_1
\widehat{C} M} +\beta \sum^{N-1}_{n=1} f^{(j)}_{n}|000\rangle_{A_1
\widehat{C} M}+ \beta \sum _{l=1} ^j f^{(l)}_N |00l\rangle_{A_1
\widetilde{C} M} \nonumber\\
\label{tt1}
 \end{eqnarray}
Then, apply a CNOT gate at the qubit of $M_1$ controlled by Bob's
site being $1$ (CNOT:$|0\rangle \langle 0 |\otimes I +|1\rangle
\langle 1 | \otimes \sigma_x$ ), the state is
\begin{eqnarray}
 |\Psi_{A_1}(t_1)\rangle
&=&  \alpha\sum^{N-1}_{n=1} f_{n}^{(1)}|0n0\rangle_{A_1
\widehat{C} M}+\beta \sum^{N-1}_{n=1} f^{(j)}_{n}|000\rangle_{A_1
\widehat{C} M} \nonumber\\
&+&\beta \sum _{l=2} ^j f^{(l)}_N |00l\rangle_{A_1
\widetilde{C} M} + f_N ^{(1)}(\alpha |N\rangle_{\widehat{C}}+ \beta |0\rangle_{\widehat{C}})
\otimes | 0\rangle_{A_2}\otimes |1\rangle_{M_1}\nonumber\\
\label{tt1}
 \end{eqnarray}
 Now, we
 measure the spin of $M_1$ . If the result is $|1\rangle$, the
 state of system is
 \begin{equation} (\alpha |N\rangle_{\widehat{C}}+ \beta | 0\rangle_{\widehat{C}})
\otimes | 0\rangle_{A_2}\otimes |1\rangle_{M_1} \end{equation} The
state of Bob's site is $|\psi\rangle = \alpha |1\rangle + \beta |
0\rangle$. Therefore we have completed the task of quantum state
transfer, and the probability is $\eta_1=|f_{N}^{(1)}|^2$. For
example, the spin chain with equal coupled strength \cite{r3}
\begin{equation} |f_{N}^{(1)}|^2=\mid\frac{2}{N+1}\sum_{k=1}^N
\sin\Big(\frac{\pi k }{N+1}\Big) \sin \Big(\frac{\pi k
N}{N+1}\Big)e^{-i E_k t}\mid ^2, ~~ E_k=-2 \cos \frac{k\pi}{N+1}.
\label{jl}
\end{equation} if the result of measurement is $|0\rangle$, the state of
system is
\begin{equation}
 \frac{1}{\sqrt{P(1)}}\{ \alpha\sum^{N-1}_{n=1} f_{n}^{(1)}|0n0\rangle_{A_1
\widehat{C} M}+\beta \sum^{N-1}_{n=1} f^{(j)}_{n}|000\rangle_{A_1
\widehat{C} M}+\beta \sum _{l=2} ^j f^{(l)}_N |00l\rangle_{A_1
\widetilde{C} M}\}
\end{equation}
where $P(1)=1-|f_{N}^{(1)}|^2$ denotes the probability of failure
for the first transfer. Generally, for randomly coupled spin
chain, $\eta_1\neq 1$.  By the measurement, we can determine
whether the state of Bob's site is $|\psi\rangle$. Because the
transfer is conclusive, and there are several measurements, it is
more convenient to use the probability of successful transmission
$\eta$ than the fidelity to measure the ability of transfer.

If the first step of decoding failed, let the spin chain evolves
freely, at time $t_2=t_1+\tau_2$, apply a CNOT gate at $M_2$
controlled by Bob's site, and measure the qubit of $M_2$. If the
result is $|1\rangle$, the state of Bob's site is $|\psi\rangle=
\alpha|1\rangle+ \beta |0\rangle$ , and the probability is
\begin{equation}
 \eta_2=|f_{N}^{(2)}|^2 =|\sum_{n=1}^{N-1}~ _{A_1
\widetilde{C} M} \langle 0N0|U(\tau_2)|0n0\rangle_{A_1 \widetilde{C}
M}\langle 0n0| U(\tau_1)|010\rangle_{A_1 \widetilde{C} M}|^2
\end{equation} If the result is $0$, the state of Bob's site is
$|0\rangle$. Then we repeat these operations at
$t_3=t_2+\tau_3$,$t_4=t_3+\tau_4,\cdots$ until the result of
measurement of $M_j$ is $|1\rangle$. We introduce $\eta_j$ to denote
the probability of successful transfer at $j$th step(the result of
$j$th measurement is $|1\rangle$ ), and the probability of
successful transfer within $j$ steps is $\eta=\sum_{i=1}^{j} \eta
_i$. Obviously, $\eta$ is an  increasing function of $j$. And it has
been proved that perfect quantum state transfer can be achieved
\cite{r19} when $j\rightarrow\infty$. So, we  conclude that when the
number of memories used $k\rightarrow\infty$,  and the number of
decoding steps $j\rightarrow\infty$,  the state can be perfectly
transferred from Alice to Bob.

\section{Numerical results and time scale}
%%%%%%%%%%%%%%%%%%%%%%%%%%%%%%%%%%%%%
\begin{figure}
  \centering
    \label{transfer:firt}
    \includegraphics[scale=1]{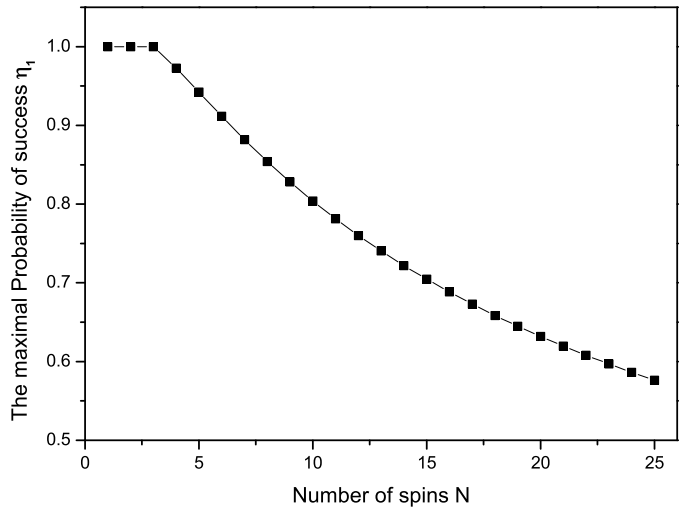}  \hspace{0.5cm}
  \caption{the maximal probability of successful transfer at the first decoding step $\eta_1$ as a function number of spins $N$ of the chain.}
  \label{transfer}
\end{figure}

%%%%%%%%%%%%%%%%%%%%%%%%%%%%%%%%%%%%%%%%%%%%%5
\begin{figure}
\centering
 \label{transfer:next}
   \includegraphics[scale=1]{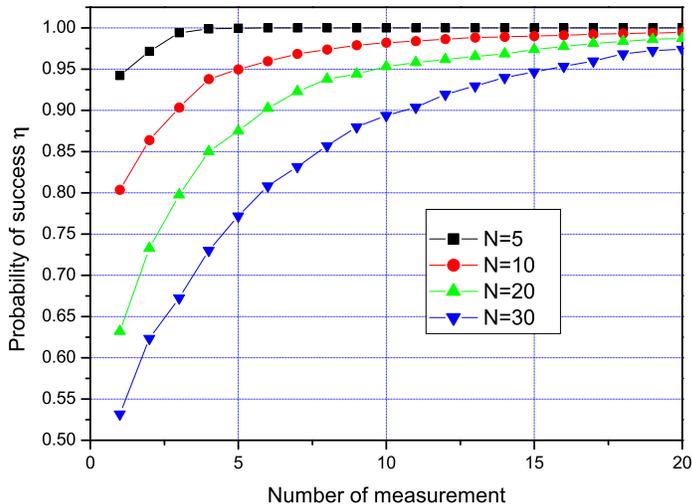}
\caption{  the probability of successful transfer within $j$
steps(measurements). The length of the spin chain is
$N=5,10,20,30$, the interval of time $\tau_i$ is chosen to make
$\eta_i$ the maximum.}
 \label{transfer2}
\end{figure}

%%%%%%%%%%%%%%%%%%%%%%%%%%%%%%%%%%%%%%%%%%%%%5
As an example, in this section, We consider the equal
nearest-neighbor coupling spin chain. The Hamiltonian of the
chain($A_1+C+B$ or $A_2+C+B$) is
 \begin{equation}
H=\frac{J}{2}
\sum_{i=1}^{N-1}\{\sigma_i^x\sigma_{i+1}^x+\sigma_i^y\sigma_{i+1}^y
\} . \end{equation} The eigenvalues and eigenvectors have been
solved exactly \cite{r3}. The probability of successful transfer
within $j$ steps $\eta = \sum_{i=1}^j \eta_i(\tau_i,N)$ is a
function of number of spins $N$ , the interval time at the $i$th
step $\tau_i$ and $j$. From Eq.(\ref{jl}), when $j=1$, and for a
certain $N$, $\tau_i$ is chosen to make $\eta_i$ achieve its
maximum, we get the maximal probability of successful transfer at
the first step, as depicted in Fig.\ref{transfer}. Because of the
dispersion of the information along the chain, transfer is not
perfect when the number of spins is larger than $3$, and the longer
the spin chain, the less the maximal probability $\eta_1$. In
Fig.\ref{transfer2}, $\tau_i$ is also chosen to make $\eta_i$
maximal. And this figure shows that when the number of measurement
$j$ (number of decoding steps) increases, the probability of
successful transfer $\eta$ approaches $1$ for all N.

The number of memories used is determined by the probability of
successful transfer $\eta$ that we requires and the transfer
ability of the spin chain. If we take  $j$ steps to transfer the
information to the memories, the time spent is $t_j=\sum_{i=1}^j
\tau_i$,~Burgarth \cite{r10} and ~Giovannetti \cite{r9} have made
a timescale estimation for $t_j$. In the decoding portion, if the
decoding protocol is Time-reverse,  the time spent on decoding is
$t_j$, because all the $j$ steps should be finished to achieve the
probability of successful transfer that we required. However, in
our scheme, once the result of measurement of memories is
$|1\rangle$, we can omit the following $(j-i)$ steps and complete
the task of transfer. Therefore, we should use the average
decoding time to measure the time for decoding.
\begin{equation} \overline{T}=\sum_{i=1}^{j-1}\eta_i
t_i+(1-\sum_{i=1}^{j-1}\eta_i)t_j \end{equation}The probability of
successful transfer within $j-1$ steps is $\sum_{i=1}^{j-1}\eta_i$,
so we have the probability $1-\sum_{i=1}^{j-1}\eta_i$ to go to  the
last decoding step. From Fig.\ref{transfer}, the maximal probability
of success is very large at the first step, i.e, when $N<25$, we
have $ \eta_1>0.5$. And from Fig.\ref{transfer2}, generally,
$\eta_i$ is the decreasing function of $i$. So, we expect that
$\overline{T}$ is obviously less than $t_j$.  For example, when
$N=10,J=20K\times \emph{k}_B$. Fifteen memories is needed to achieve
the probability of successful transfer $\eta=0.99$. The time spent
on the transfer portion is $t_{15}=0.37ns$, which is also the time
spent on the decoding portion if we use the Time-reversed protocol.
However, in our scheme, the average decoding time is
$\overline{T}=3.02\times10^{-2}ns$, which is only
$\frac{1}{12}t_{15}$.

\section{Conclusion}
Because the information can be easily destroyed by measurement,
conclusive transfer in a single spin chain was once thought to be
unable \cite{r11}. Nevertheless, in our scheme, we successfully
make it by using local memories storing the information and
dividing the task into transfer portion and decoding portion. When
the number of memories used and the decoding steps are infinite,
our scheme can get perfect state transfer. Especially, our scheme
is much better than Time-reversed protocol in some aspects. First,
our scheme do not need another identical spin chain. Second,
considering the timescale, the average decoding time in our scheme
is much less than the Time-reversed protocol. Third, our scheme is
conclusive transfer.

\vspace{.5cm} {\bf Acknowledgments}:

The work was supported  by the Fund of Chinese Academy of
Sciences, the Education Ministry of China, and the National
Natural Science Foundation of China under Grant No 10231050.

\newpage

\end{document}